\documentclass[preprint]{aastex63}
\usepackage{amssymb,amsmath}
\usepackage{graphicx}
\usepackage{natbib}
\usepackage[OT1,T1]{fontenc}

\begin{document}

\title{Modeling the IRIS Lines During a Flare. I. The Blue-Wing Enhancement in the \ion{Mg}{2} k Line}

\author{Jie Hong}
\affiliation{School of Astronomy and Space Science, Nanjing University, Nanjing 210023, China}
\affiliation{Key Laboratory for Modern Astronomy and Astrophysics (Nanjing University), Ministry of Education, Nanjing 210023, China}

\author{Ying Li}
\affiliation{Key Laboratory of Dark Matter and Space Astronomy, Purple Mountain Observatory, Chinese Academy of Sciences, Nanjing 210034, China}

\author{M.~D. Ding}
\affiliation{School of Astronomy and Space Science, Nanjing University, Nanjing 210023, China}
\affiliation{Key Laboratory for Modern Astronomy and Astrophysics (Nanjing University), Ministry of Education, Nanjing 210023, China}

\author{Yu-Hao Zhou}
\affiliation{Centre for mathematical Plasma Astrophysics (CmPA), KU Leuven, Celestijnenlaan 200B, B-3001 Leuven, Belgium}
\affiliation{School of Astronomy and Space Science, Nanjing University, Nanjing 210023, China}

\email{jiehong@nju.edu.cn}

\begin{abstract}
The IRIS \ion{Mg}{2} k line serves as a very good tool to diagnose the heating processes in solar flares. Recent studies have shown that apart from the usual red asymmetries which are interpreted as the result of condensation downflows, this line could also show a blue-wing enhancement. To investigate how such a blue asymmetry is formed, we perform a grid of radiative hydrodynamic simulations and calculate the corresponding line profiles. We find that such a spectral feature is likely to originate from the upward plasma motion in the upper chromosphere. However, the formation region that is responsible for the blue-wing enhancement could be located in an evaporation region, in an upward moving blob, and even an upward moving condensation region. 
We discuss how the electron beam parameters affect these different dynamics of the atmosphere.
\end{abstract}


\section{Introduction}
The \ion{Mg}{2} k and h lines, which originate from the transition between the \ion{Mg}{2} $3s$ and $3p$ terms, are the strongest lines in the IRIS \citep{2014depontieu} near-ultraviolet passband. These resonance lines are usually optically thick and have a wide range of formation height, from the mid to upper chromosphere \citep{2013leenaartsb}. Thus these lines are good indicators of solar activities that have a large influence on the chromosphere, especially solar flares. 

The thick-target model of solar flares suggests that the released energy in the solar corona would accelerate a large amount of non-thermal electrons. These electrons then propagate downwards along flare loops, and deposit their energy in the chromosphere through Coulomb collisions. The heating of the chromospheric plasma is strongly enough that the local temperature reaches the coronal temperature. This process is defined as the chromospheric evaporation \citep{1985fishera}, and the heated material usually expands upward into corona as a consequence. \cite{1985fisherb} noticed that there are two kinds of chromospheric evaporation. For weaker flares, the evaporation is gentle; while for strong flares the evaporation would become explosive. Explosive evaporation is always accompanied with a downward-moving compression wave. Behind this wave is a region with cooler and denser plasma compared with the adjacent chromospheric plasma. This region is referred to as the chromospheric condensation \citep{1985fisherc}. In most situations, the condensation region moves downwards as a result of the compression wave above \citep{1985fisherc,1989fisher}. The plasma motion that is connected to flare dynamics has long been observed and simulated through Doppler shifts or line asymmetries of various spectral lines \citep{1984ichimoto,1999abbett,2004brosius,2015kuridze}.

In solar flares, the observed \ion{Mg}{2} k line is greatly enhanced and broadened, but with very diverse shapes \citep{2017rubio,2018panos}. Usually, this line is redshifted or shows an asymmetric red component \citep{2015kerr,2017rubio,2019li}. The red asymmetry of this line indicates downflows in the chromosphere, which possibly indicates chromospheric condensations \citep{2019li}. However, despite the usual red asymmetry, this line could also show a blue wing enhancement at the flare footpoints \citep{2015cheng,2015kerr,2018tei,2019huang,2019li}. A straightforward explanation of the blue-wing enhancement would be plasma upflows which result from the chromospheric evaporation \citep{2015cheng,2019li}. 
The simulations from \cite{2016kerr} have confirmed that the blue-wing emissions are enhanced by upflows.
\cite{2015kerr} argued that besides the probability of upflows, such a blue asymmetry can also be due to absorptive plasma with a downward velocity. \cite{2018tei} reported a similar profile shape in the \ion{Mg}{2} h line, but proposed a new scenario regarding the upflows. They thought that the upflows do not result from the chromospheric evaporation, but represent some cold plasma which is pushed upwards by the expansion of the hot plasma below. \cite{2019huang} made the first trial to interpret such a blue asymmetry from the simulation results, and found an increase of electron density and velocity field in the \ion{Mg}{2} line formation region. However, it is still not very clear whether the proposed scenario is a feasible explanation. In particular, the questions as to what conditions are required for the origin of such an upflow, and how the line is formed during this process, need to be studied further.

In this paper, we use radiative hydrodynamic simulations to study the \ion{Mg}{2} k line in response to heating in a flare. We investigate the physical nature behind the appearance of the blue wing enhancement and examine the relevant scenarios. The rest of the paper is organized as follows. We introduce our method in Section~2, and present the results in Section~3. In Section~4 we make discussions and draw our main conclusions.

\section{Method}
\begin{figure}
\plotone{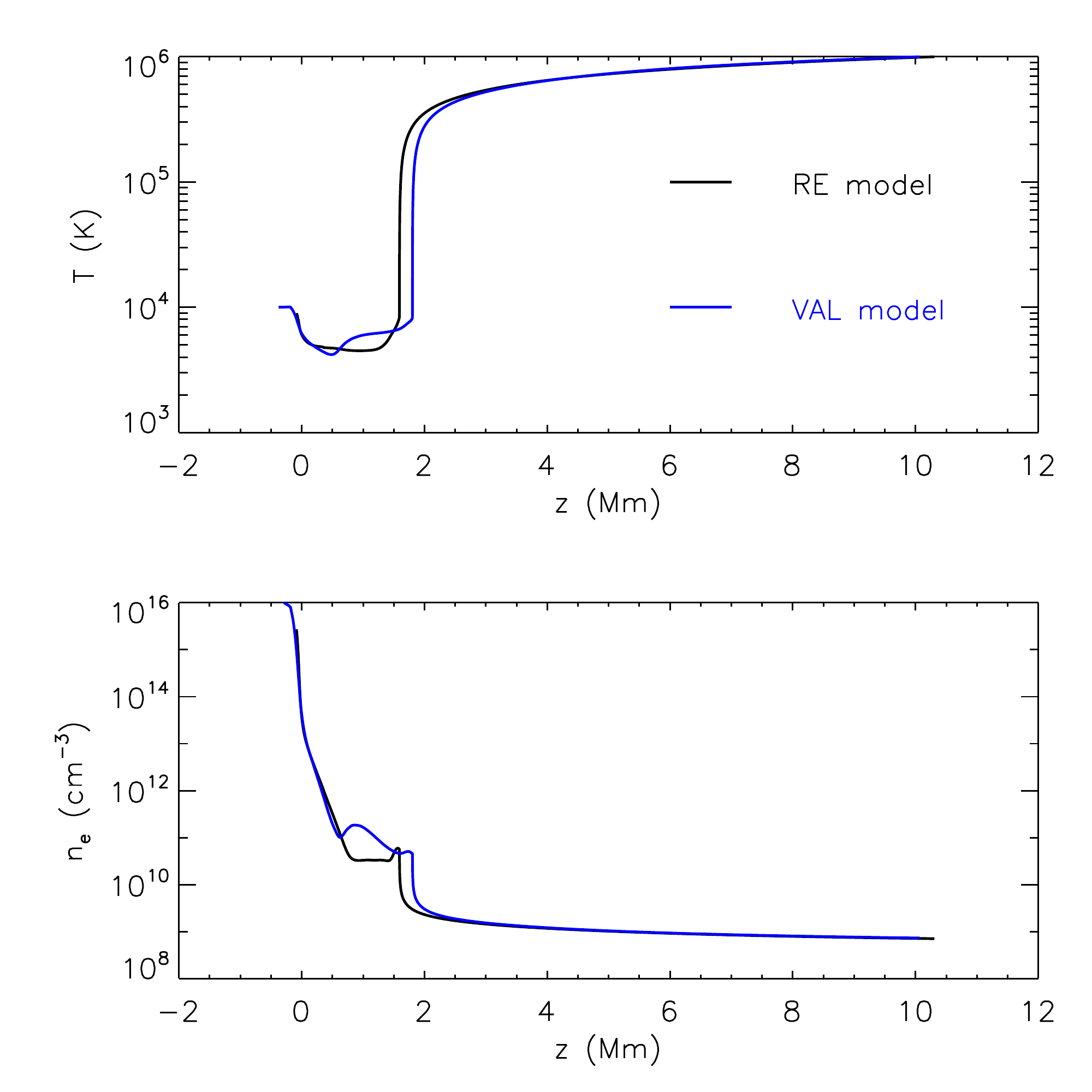}
\caption{Height distribution of temperature and electron density of our VAL model and the radiative equilibrium (RE) model in \cite{2015allred}. }
\label{atm}
\end{figure}

We use the 1D radiative hydrodynamics code RADYN \citep{1992carlsson,1995carlsson,1997carlsson,2002carlsson} to calculate the atmospheric response to a beam of non-thermal electrons injected at the loop top during a flare. RADYN solves the hydrodynamic equations in combination with the non-LTE radiative transfer equation and the population rate equation along a pre-defined loop structure. The loop has a quarter-circular shape with a 10 Mm length. 
The initial atmosphere is constructed by adding a corona to the VAL model \citep{1981vernazza} and then relaxing the whole atmosphere with a fixed temperature of 1 MK at the loop top and an artificial heating in the lower atmosphere to keep the temperature structure \citep{2017hong,2018hong,2019hong}. Our pre-flare atmosphere is slightly different from the radiative equilibrium atmospheric models in \cite{2015allred}. The main difference is that in our model, there is a chromosphere at a height of around 1 Mm with increased temperature and electron density (Fig.~\ref{atm}). The transition region is also higher in our model.
The injected non-thermal electron beam is assumed to have a power-law distribution with a total energy flux $F$, a cut-off energy $E_{c}$ and a spectral index $\delta$. 
We use the Fokker-Planck approach that has been implemented in RADYN to calculate the heating rate from the non-thermal electron beam \citep{2015allred}, but we did not include any return current.
 The upper boundary is set to be reflective to account for the symmetry of the flare loop. We have run a grid of flare simulations, and show their parameters in Table~\ref{tab}. We label each simulation in the form of f$n_1$E$n_2$d$n_3$, where the numbers $n_1$ to $n_3$ stand for the values of $\log F_{peak}$, $E_c$ and $\delta$, respectively. In the first six models, the beam heating function rises linearly with time, while in the last two models, the heating stays constant.

\begin{table}[ht]
\centering
\caption{List of parameters of the flare models for simulation}
\begin{tabular}{cccccc}
\hline
Label & $F_{peak}$ (erg cm$^{-2}$ s$^{-1}$) & Heating function & Total duration (s) & $E_{c}$ (keV)  & Spectral index\\
\hline
f10E20d4 & $10^{10}$ & linear & 10  & 20 & 4 \\
f10E15d5 & $10^{10}$ & linear & 10  & 15 & 5 \\
f10E10d6 & $10^{10}$ & linear & 10  & 10 & 6 \\
f11E20d4 & $10^{11}$ & linear & 10  & 20 & 4 \\
f11E15d5 & $10^{11}$ & linear & 10  & 15 & 5 \\
f11E10d6 & $10^{11}$ & linear & 10  & 10 & 6 \\
f9E20d7 & $10^{9}$ & constant & 20  & 20 & 7 \\
f10E20d7 & $10^{10}$ & constant & 20  & 20 & 7 \\
\hline
\end{tabular}
\label{tab}
\end{table}

Similar to the Ly$\alpha$ line, the \ion{Mg}{2} resonance lines are greatly influenced by partial frequency redistribution (PRD) \citep{2019kerra}. As RADYN assumes complete frequency redistribution (CRD) for all spectral lines, we use the RH code \citep{2001uitenbroek,2015pereira} to take account of the PRD effect. We save the snapshots of the RADYN simulations every 0.1 s and feed the RADYN outputs into RH to calculate the \ion{Mg}{2} k line. We use a 10-level-plus-continuum \ion{Mg}{2} atom that is the same as in \citet{2013leenaarts}. Nevertheless, RH solves the radiative transfer equation under the assumption of statistical equilibrium instead of non-equilibrium ionization, which can change the ionization fraction and lead to a different line formation height \citep{2019kerrb}. In order to mitigate this effect, we adopt the values of the electron density and hydrogen density from RADYN that conform to the non-equilibrium ionization and fix them in RH calculations, as done in \cite{2017rubio} and \cite{2019kerra}. 
We do note that for those weaker flares in our simulations (the f9 and f10 models), there will be non-equilibrium effects in the main heating phase \citep{2019kerrb}.
However, for strong flares, the difference in the \ion{Mg}{2} resonance line profiles between statistical equilibrium and non-equilibrium ionization is smaller than the difference between PRD and CRD  in the main heating phase \citep{2019kerrb}.

\section{Results}
\begin{figure}
\plotone{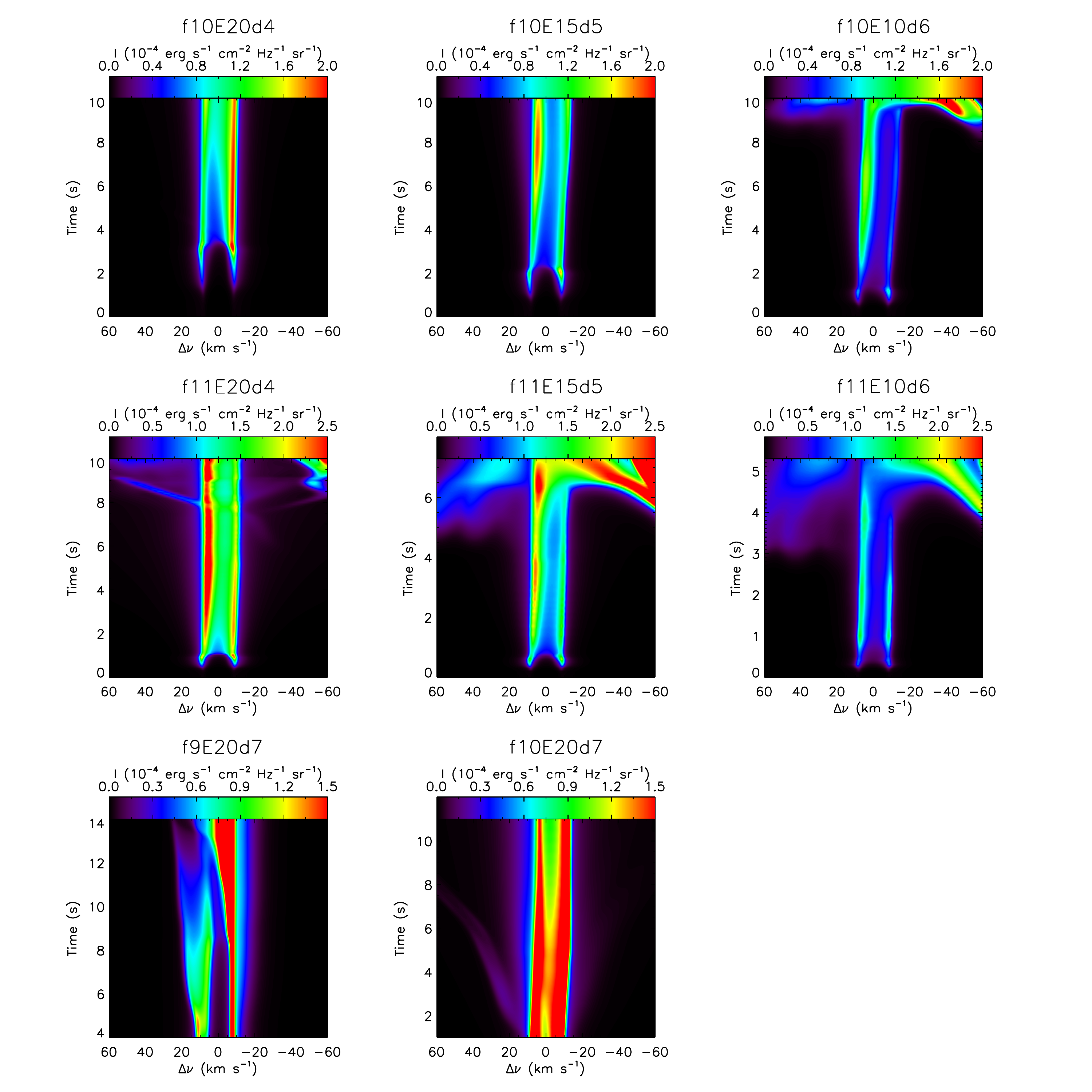}
\caption{Time evolution of the \ion{Mg}{2} k line profiles in the eight flare models. The wavelength in each panel is in Doppler scale, with positive values for the blue wing and negative values for the red wing.}
\label{int}
\end{figure}

The time evolution of the \ion{Mg}{2} k line profiles of all the flare models are shown in Fig.~\ref{int}. One can see clearly that the $k_{2}$ peaks are greatly enhanced when heating begins. The line profiles can show either a red asymmetry (a larger $k_{2r}$ peak) or a blue asymmetry (a larger $k_{2v}$ peak), as a result of the mass flows in the region where the line core is formed \citep{2013leenaartsb}. The line profiles become more complicated when heating continues. In some models (e.g. models f11E20d4, f11E15d5 and f11E10d6), an intensity component appears in the red wing that corresponds to the chromospheric condensation. More interestingly, in three models (models f11E20d4, f9E20d7 and f10E20d7), an intensity component moves from the line core to the blue wing during the heating phase, showing a very obvious blue-wing enhancement in the line profiles. Below we describe these three models in detail.

\subsection{Model f11E20d4}

The most obvious blue-wing component in the \ion{Mg}{2} k line appears in the f11E20d4 model. We show the atmospheric structure and line formation for this model at 8.2 s in Fig.~\ref{104}. At this time, the atmosphere is experiencing an explosive chromospheric evaporation, with a velocity of 100 km s$^{-1}$ and a temperature of $3\times10^{6}$ K at 1.8 Mm (Fig.~\ref{104}(a) and (b)). We notice that there appears to be two obvious condensation regions in the atmosphere, which seems a common result in the simulations \citep{2015kennedy,2019hong}. We show in Fig.~\ref{cond} how these condensation regions are formed. When the chromosphere gets heated continuously, the temperature rises and the first condensation region appears at a height of 1.6 Mm at 4.0 s. This region is cooler than the nearby regions and it moves downwards with a velocity of 40 km s$^{-1}$. Then the second condensation region appears at a lower height of 1.4 Mm at 6.5 s. After that, the downward motion of the first condensation region (which is somewhere above the second one) is decelerated, and it eventually moves upwards, showing a C shape in the time-height diagram. 

A zoom-in of the upper condensation region at around 1.51 Mm is shown in Fig.~\ref{detail1}. As seen in Fig.~\ref{104}(a) and Fig.~\ref{detail1}(a), the local plasma in the two condensation regions is relatively cooler, and has a temperature of $\sim2$--$3\times 10^4$ K (Fig.~\ref{detail1}(a)). In between the two condensation regions, the plasma is still hot and over $10^6$ K, thus experiencing an expansion there. This produces a nearly linear distribution of the velocity along height between the two condensation regions (Fig.~\ref{detail1}(b)). The lower condensation region has a downward velocity of $-50$ km s$^{-1}$ and the upper condensation region has an upward velocity of 20 km s$^{-1}$ (Fig.~\ref{104}(b)). The \ion{Mg}{2} density in the upper condensation region is about four orders of magnitude larger than in the nearby regions (Fig.~\ref{104}(e) and Fig.~\ref{detail1}(d)). Thus, the opacity is significantly increased in the upper condensation region and a large fraction of the emergent intensity originates there. 

The contribution function to the emergent intensity is defined as $C_I\equiv dI/dz=j_\nu \mathrm{e}^{-\tau_\nu}$, if seen vertically, where $j_\nu$ is the emissivity and $\tau_\nu$ is the optical depth.
We show the contribution function $C_I$ at two wavelengths in Fig.~\ref{104}(c) and (d). The mean formation height, defined as $z_h=\int zC_{I} dz/\int C_{I} dz$ \citep{2012leenaarts}, is denoted with a red vertical line. 
This quantity can better evaluate in which region the line is formed compared with the height where $\tau=1$, especially when there are multiple peaks in the contribution function.
It is very clear that the condensation regions have a large influence on the contribution function. In the quiet Sun, the blue wing at $\Delta\nu = 20$ km s $^{-1}$ is usually formed in the lower chromosphere. In a flaring atmosphere, although there is still a large contribution from the lower chromosphere at a height of $\sim$0.9 Mm, the contribution from the upper condensation region at a height of $\sim$1.5 Mm is largely enhanced and the value of $C_I$ gets more than three orders of magnitude larger. As a result, the mean formation height is shifted to the upper chromosphere. In the line core, however, the contribution from the condensation regions is not very large. The line core is still formed in the mid-chromosphere as in the quiet Sun. We further integrate the contribution function $C_{I}$ over the layers between 1.50 Mm to 1.52 Mm, which cover the whole upper condensation region (see the blue shades in Fig.~\ref{detail1}(a)--(d)). The integration results in a sole blue-wing component in the line profile (Fig.~\ref{detail1}(e)), which has a peak amounting to more than 80\% of the emergent intensity at $\Delta\nu = 20$ km s $^{-1}$ (Fig.~\ref{detail1}(f)). The emergent line profile, as shown in Fig.~\ref{104}(f), is then very interesting. Apart from the double-peaked line core, there are another two components located in the blue wing and the red wing, respectively. The blue-wing component is at $\Delta\nu = 20$ km s $^{-1}$, which corresponds to the upflow in the upper condensation region, while the red-wing component is at $\Delta\nu = -55$ km s $^{-1}$, which corresponds to the downflow in the lower condensation region (Fig.~\ref{cont}). 
The red-wing component that originates from condensation downflows is a very common feature in flare simulations \citep{2019kerrsi,2019kerra}.

\subsection{Model f9E20d7}

We then show another case with an obvious blue-wing enhancement. Fig.~\ref{97} and \ref{detail2} show the atmospheric structure and line formation for model f9E20d7 at 8.8 s. At this time, the chromosphere has been continuously heated and the temperature has been raised to more than $10^4$ K (Fig.~\ref{97}(a)). Heating of the chromosphere drives the hot chromospheric plasma to move upwards, which appears to be the scenario of gentle evaporation (Fig.~\ref{97}(b)).
\cite{2015allred} has already shown that when calculating the beam heating rate through the Fokker-Planck approach, the results would show a tail stretching to a larger height in the atmosphere if the spectral index of the beam spectrum is smaller.
On the other hand, if the beam spectrum is very soft with a large spectral index, as is the case with model f9E20d7, the tail would quickly reach zero at a relatively low height and keeps the upper layers (from 1.70 Mm to 1.80 Mm) of the atmosphere nearly undisturbed, as can be seen from the beam heating profiles in Figs.\ref{97}(a), \ref{107}(a) and \ref{blob}, and the Fig. 2(n) of \cite{2018polito}. We then focus on the upper chromosphere and find that the heated layers located below are pushing the unheated cool layers upwards, with a velocity of around 13 km s$^{-1}$ at the boundary (Fig.~\ref{detail2}(b)). However, different from the previous model f11E20d4, we do not find any large spikes in the \ion{Mg}{2} density distribution in the upper chromosphere (Fig.~\ref{97}(e) and Fig.~\ref{detail2}(d)). Since there is an upward velocity, the contribution function at the blue wing ($\Delta\nu = 16$ km s $^{-1}$) in this region is four orders of magnitude larger than that in the nearby regions, and the formation height is also shifted to this region (Fig.~\ref{97}(c) and Fig.~\ref{detail2}(c)). We again investigate the contribution to the line intensity of the evaporating plasma in the height range from 1.64 Mm to 1.70 Mm (blue shades in Fig.~\ref{detail2}) and the cool region from 1.70 Mm to 1.80 Mm (orange shades in Fig.~\ref{detail2}). We find that the blue wing enhancement is solely due to the evaporation upflow, while the nearly undisturbed cool region above it only contributes to the line core intensity. The line profile then  has an obvious blue-wing enhancement, just at the blue side of the $k_{2v}$ peak. 

\subsection{Model f10E20d7}

Model f10E20d7 has almost the same set of parameters as model f9E20d7, except for a larger electron beam flux. We show the atmospheric structure and line formation for this model at 3.0 s in Fig.~\ref{107} and \ref{detail3}. In this case, the chromosphere is heated to a higher temperature due to the larger electron beam flux, yet the upper layers from 1.76 Mm to 1.80 Mm still remains nearly undisturbed. However, what is new is that a blob appears at the bottom of the cool region, which is formed as a result of energy imbalance. As shown in Fig.~\ref{blob}, work done by pressure gradient and viscous heating supply much energy to this region since the plasma below is continuously moving upwards. As this energy cannot be radiated away promptly, the temperature of the blob is then slightly enhanced to $\sim1$--$2\times10^4$ K (Fig.~\ref{detail3}(a)), but still much lower than that of the evaporating plasma below. The upward velocity reaches its maximum value of 25 km s$^{-1}$ at the interface of the evaporating chromospheric plasma and the cool blob (Fig.~\ref{detail3}(b)). The density in the cool blob and the undisturbed cool layers above has  increased to four orders of magnitude higher than that in the nearby regions (Fig.~\ref{107}(e) and Fig.~\ref{detail3}(d)).

The cool blob dominates in the contribution to the blue-wing intensity (Fig.~\ref{detail3}(c)). The upper cool region, however, makes a great contribution to the line core intensity, although it is still one order of magnitude less than the contribution from the mid-chromosphere (Fig.~\ref{107}(d)). The formation height of the blue wing is again shifted upwards to the cool layers while the formation height of the line core remains unshifted (Fig.~\ref{107}(c) and (d)). The line profile has a blue-wing enhancement, but less obvious than in model f9E20d7 (Fig.~\ref{107}(f)). In Fig.~\ref{detail3}(e) and (f), we plot the contribution to the line intensity from the evaporating plasma (1.64--1.735 Mm, blue shades), the blob (1.735--1.76 Mm, green shades) and the upper cool region (1.76--1.80 Mm, orange shades). One can clearly see that the blob is mostly responsible for the blue-wing enhancement, while the evaporation upflow has nearly no contribution simply because the temperature there is too high.
 
 \section{Discussion and conclusion}
We have calculated the \ion{Mg}{2} k line for a grid of flare models subject to electron beam heating with different sets of beam parameters. We find that apart from the red-wing enhancement that frequently appears in observations as a result of chromospheric condensation, a blue-wing enhancement also appears in the heating phase of some models. Generally speaking, this spectral feature originates from an upflow in the upper chromosphere. However, the location for the upflow could be different in different flare models. It can be an upward moving chromospheric condensation region (f11E20d4), a chromospheric evaporation region (f9E20d7) or an upward moving blob pushed by the evaporation below (f10E20d7). In all cases, the upflows have a temperature of about $\sim1$--$3\times10^4$ K, which is sensitive to the \ion{Mg}{2} k line. We would also like to point out that a condensation region does not necessarily move downwards \citep{2015kennedy}, as is the case with f11E20d4.

\cite{2019kerra} have noted that the \ion{Mg}{2} k line wing enhancement could be the result of the PRD effects, rather than real mass flows in the atmosphere. To check this possibility, we have carefully studied how the line profiles are formed in terms of the distributions of the source function and contribution function (Fig.~\ref{cont}). 
We find that the distribution of the source function is greatly influenced by the local velocity rather than by the PRD effects, especially at the height where the contribution function is large.
We then conclude that the blue-wing enhancement in our models is indeed caused by an upward motion of the chromospheric plasma, while the PRD effects play a minor role. 

It is always a challenge to relate line-wing enhancements or line asymmetries to the absolute velocities of mass flows in the atmosphere, especially for optically thick lines. Previous studies of the H$\alpha$ and the Ly$\alpha$ lines show that a blue asymmetry could be interpreted as condensation downflows \citep{2015kuridze,2018brown,2019hong}. 
As for the \ion{Mg}{2} k line, \cite{2013leenaartsb} showed that for the quiet Sun, a stronger blue peak implies downflows above the peak formation height. \cite{2016kerr} found that in flares, the plasma flows could result in emission peaks that appear in the \ion{Mg}{2} k line wings. Similarly, in our results, the blue-wing enhancement in the \ion{Mg}{2} k line reflects upflows in the chromosphere, although our beam parameters are different from that in \cite{2016kerr}.

However, we find different scenarios for the origin of upflows that are responsible for the blue-wing enhancement. The result from model f9E20d7 shows that the chromospheric evaporation could contribute to the blue-wing enhancement, as long as the temperature there is not raised too much to exceed the formation temperature of the \ion{Mg}{2} k line \citep{2019li}. On the other hand, a cool-upflow scenario has also been proposed by \cite{2018tei} to interpret such a blue-wing enhancement in the \ion{Mg}{2} h line. Our simulations confirm that such a physical process is feasible if the electron beam penetrates relatively deep (with a relatively high cut-off energy), thus making the upper chromosphere less disturbed during the heating phase, as is the case with f9E20d7 and f10E20d7. In such cases, the blue-wing enhancement of the \ion{Mg}{2} k line mainly originates from the evaporating plasma or the upward-moving blob just below the cool region. The cool region itself has less contribution to the blue-wing enhancement, since its velocity is relatively small. In addition, we find a third scenario as shown in model f11E20d4, in which there appears to be two condensation regions in the atmosphere, and the upper one could move upwards after being pushed by the expanding plasma below, thus contributing to the blue-wing intensity of the \ion{Mg}{2} k line. In this case, a red-wing counterpart would appear in the line at the same time  that corresponds to the lower condensation region with a downward velocity. At present, we lack observational evidence for this third scenario, although it has been tested in a number of 
 1D radiative hydrodynamic simulations.

Our grid of models also reveals the influence on the line profile shape of varying  the flare heating parameters. In particular, the low energy cut-off of the electron beam seems to relate to the appearance of the blue-wing enhancement. A similar role is for the spectral index, which determines how the beam energy is deposited along height. In the case of a very soft beam, model f10E20d7, there is almost no heating above 1.6 Mm at 1.0 s, while in the case of a relatively hard beam, model f11E20d4, the heating profile has a tail that extends to 1.75 Mm at 1.0 s. The energy flux of the electron beam is of course an important parameter in describing the heating strength. Previous studies have confirmed that chromospheric condensation only appears when the beam flux is large enough, as revealed by the different results from models f10E20d4 and f11E20d4. In f10E20d4, there is no condensation region, and the whole \ion{Mg}{2} k line is formed below the height of 1.3 Mm. 

\begin{figure}
\plotone{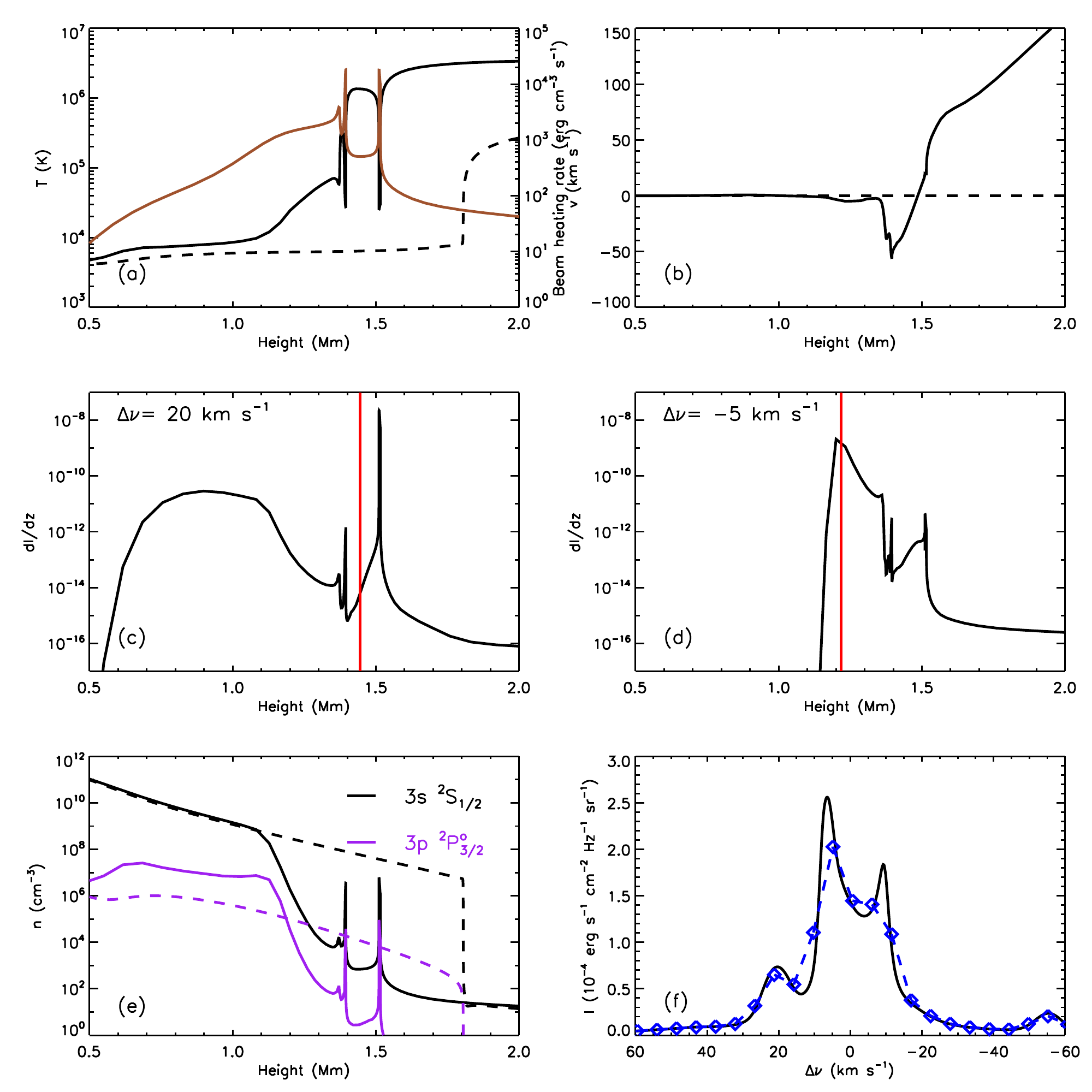}
\caption{Atmospheric structure and the \ion{Mg}{2} k line formation for the f11E20d4 model at 8.2 s. Dashed lines are for pre-flare values at 0.0 s. (a) Height distribution of the temperature (black) and the beam heating rate (brown). (b) Height distribution of the velocity, with a positive sign for upflows, and a negative sign for downflows. (c) and (d) Height distribution of the contribution function at the blue wing ($\Delta\nu = 20$ km s $^{-1}$) and that at the line core ($\Delta\nu = -5$ km s $^{-1}$). The red vertical line in each panel denotes the mean line formation height. (e) Height distribution of the number density at the lower (black) and upper (purple) levels of the \ion{Mg}{2} k line. (f) The synthetic profile of the \ion{Mg}{2} k line. The wavelength is in Doppler scale, with positive values for the blue wing and negative values for the red wing. Blue diamonds denote the synthetic profile degraded to IRIS resolution of 52 m\r{A} pixel$^{-1}$.} 
\label{104}
\end{figure}

\begin{figure}
\plotone{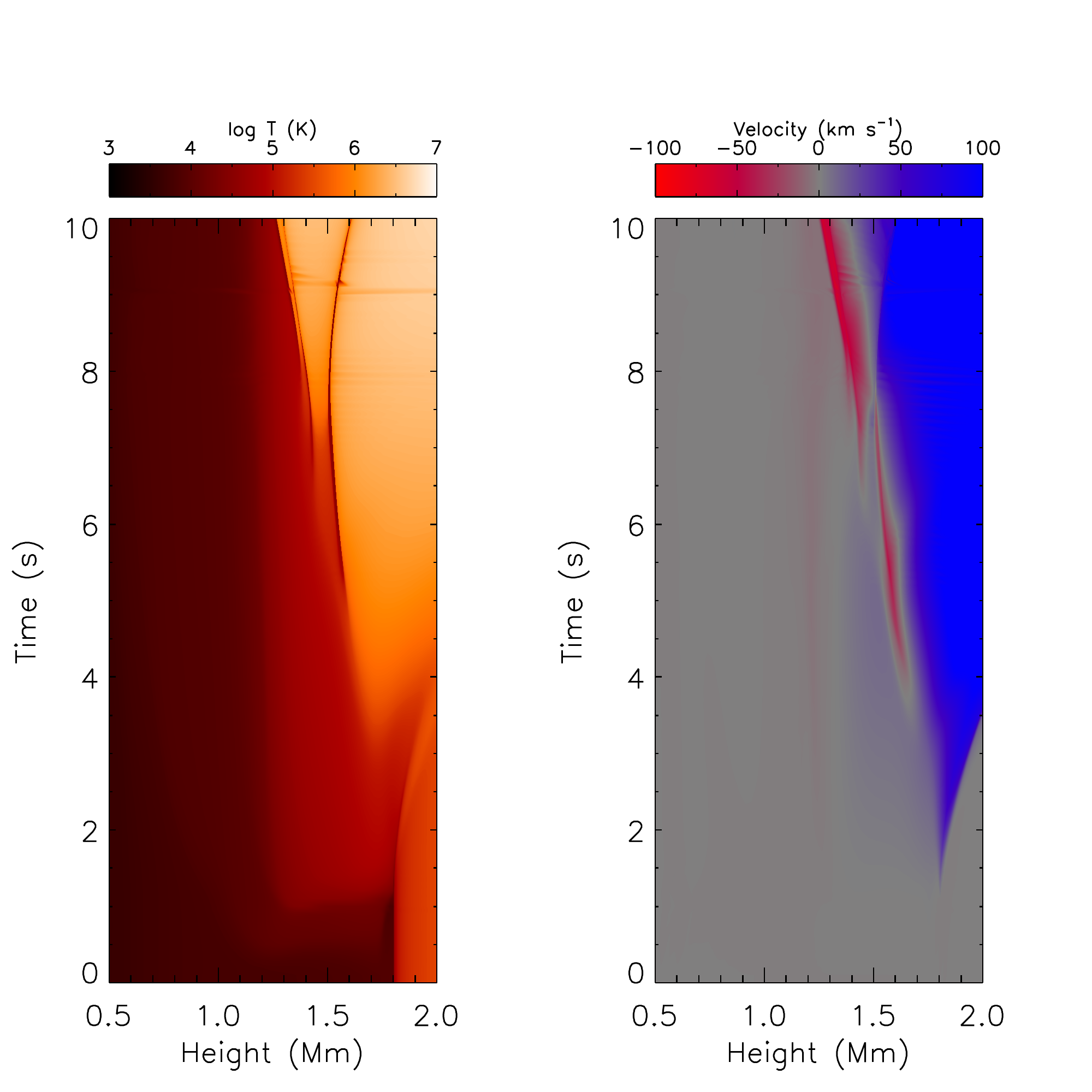}
\caption{Time-height plot of temperature (left) and vertical velocity (right) for the f11E20d4 model.}
\label{cond}
\end{figure}

\begin{figure}
\plotone{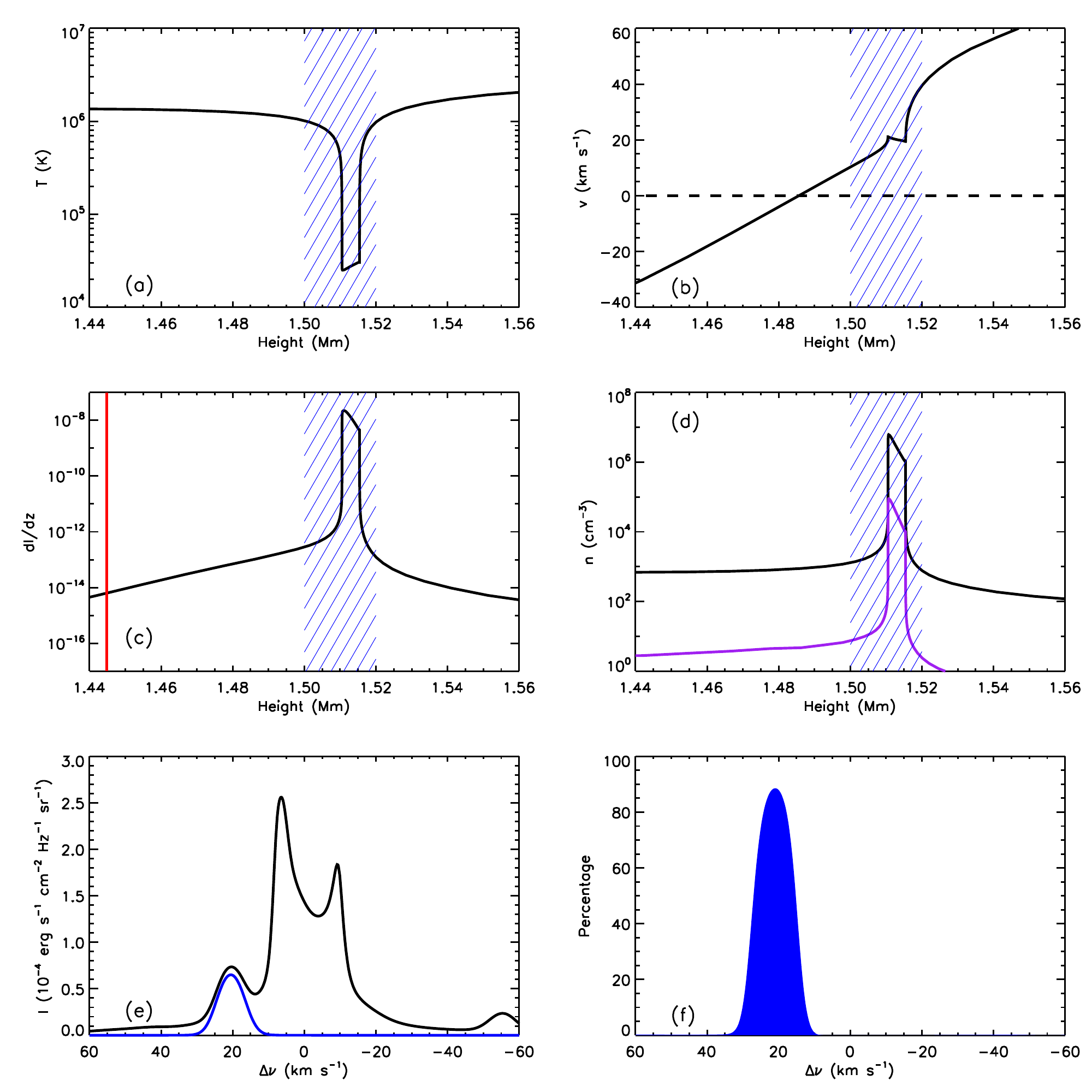}
\caption{(a)--(d) Height distribution of temperature, velocity, contribution function at the blue wing ($\Delta\nu = 20$ km s $^{-1}$) and \ion{Mg}{2} number density for the f11E20d4 model at 8.2 s. These are zoom-in plots of panels (a)--(c) and (e) in Figure~\ref{104}, highlighting the region where the blue-wing enhancement is formed. (e) The synthetic profile of the \ion{Mg}{2} k line (black). The blue line represents the line intensity contributed by the region that is shaded in panels (a)--(d). (f) The percentage of line intensity contributed by the shaded region.}
\label{detail1}
\end{figure}

\begin{figure}
\plotone{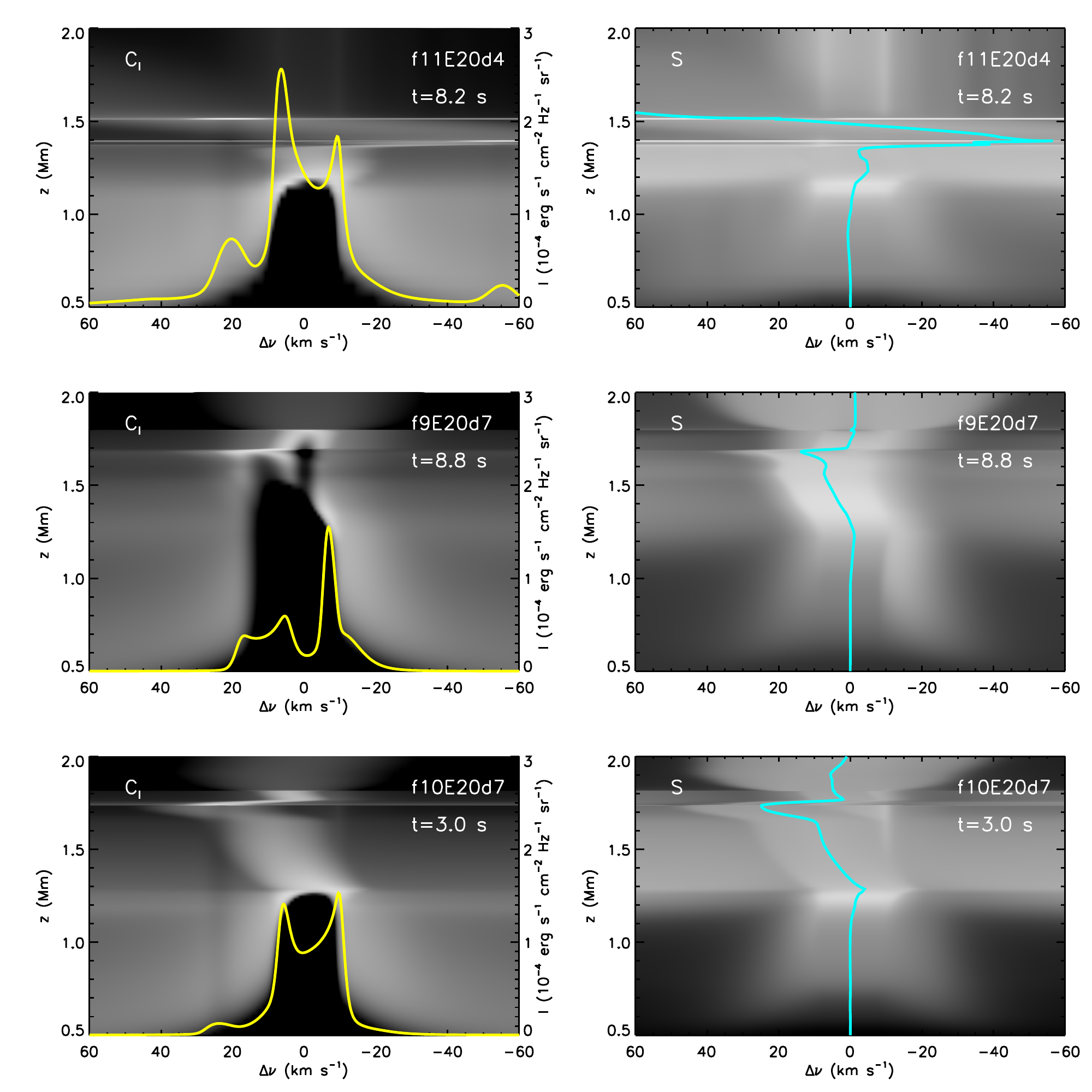}
\caption{The background gray-scale shades show the contribution functions and source functions of the \ion{Mg}{2} k line for the three models. Brighter regions indicate a larger value. The yellow lines show the \ion{Mg}{2} k line profiles, and the cyan lines denote the vertical velocity.}
\label{cont}
\end{figure}

\begin{figure}
\plotone{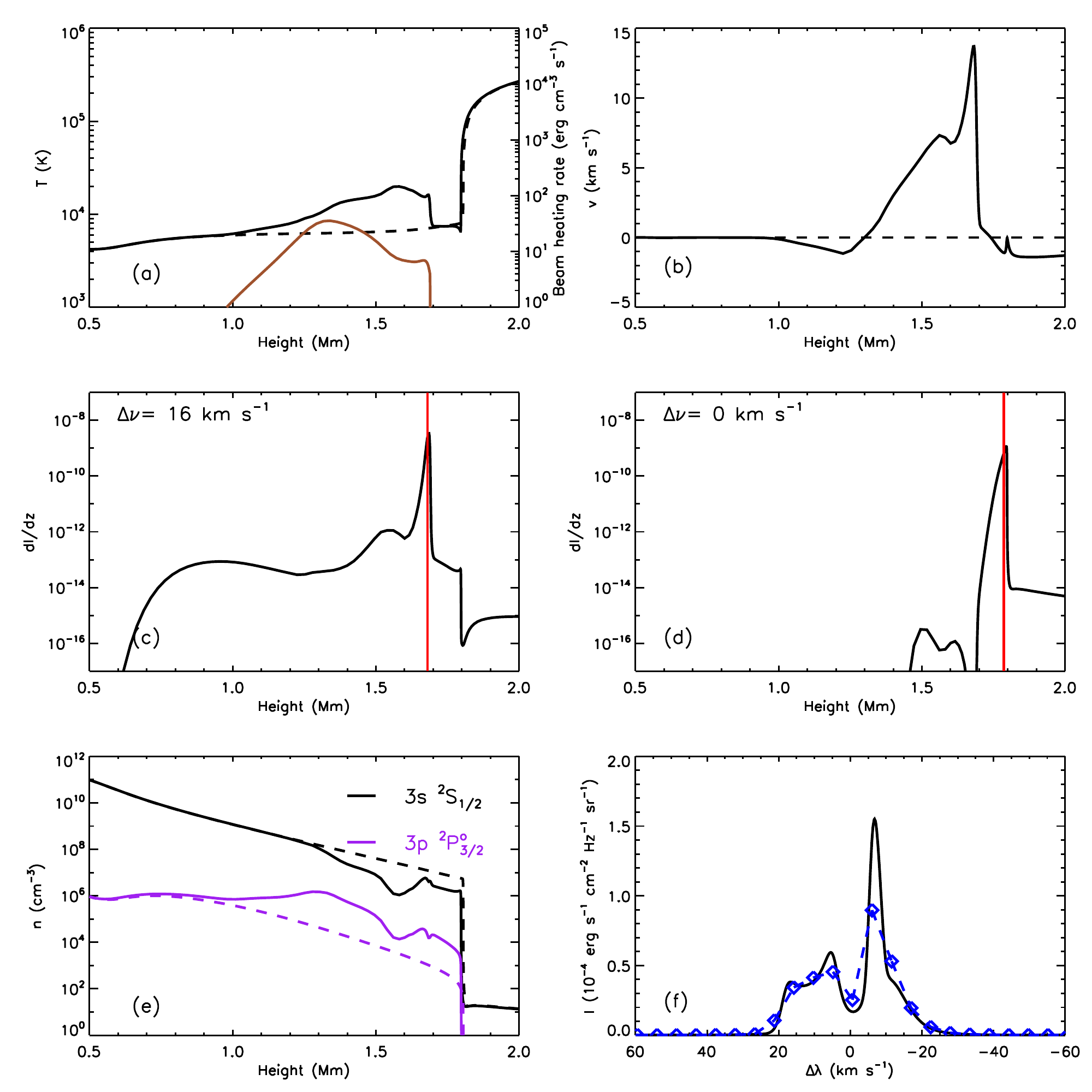}
\caption{Same as Figure~\ref{104}, but for the f9E20d7 model at 8.8 s. Note that the contribution functions in panels (c) and (d) are for $\Delta\nu = 16$ km s $^{-1}$ and $\Delta\nu = 0$ km s $^{-1}$, respectively.}
\label{97}
\end{figure}

\begin{figure}
\plotone{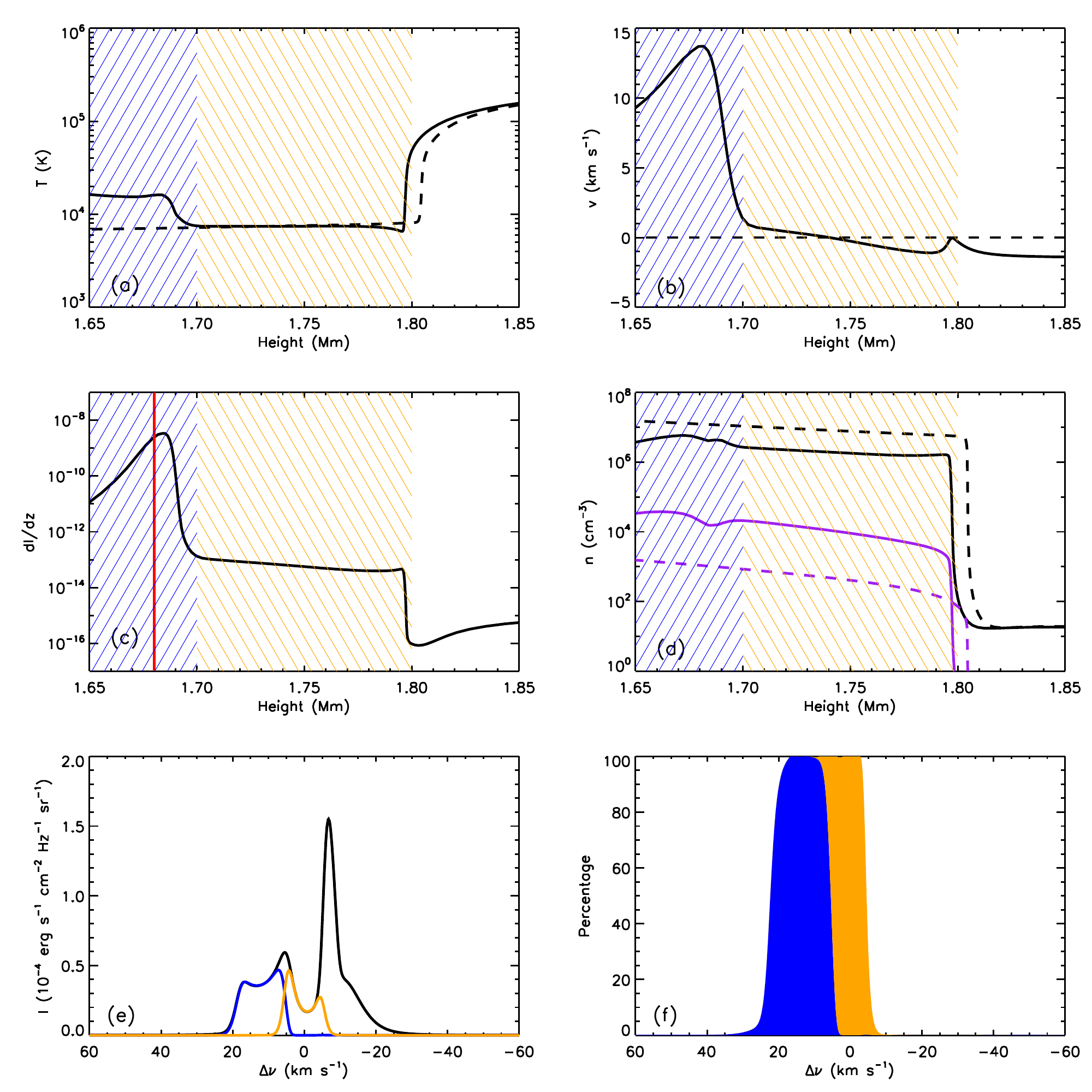}
\caption{Same as Figure~\ref{detail1}, but for the f9E20d7 model at 8.8 s. Note that the contribution function in panel (c) is for $\Delta\nu = 16$ km s $^{-1}$. The blue and orange lines in panel (e) and areas in panel (f) represent the values and percentage of line intensity contributed by the region that is shaded in the same color in panels (a)--(d).}
\label{detail2}
\end{figure}

\begin{figure}
\plotone{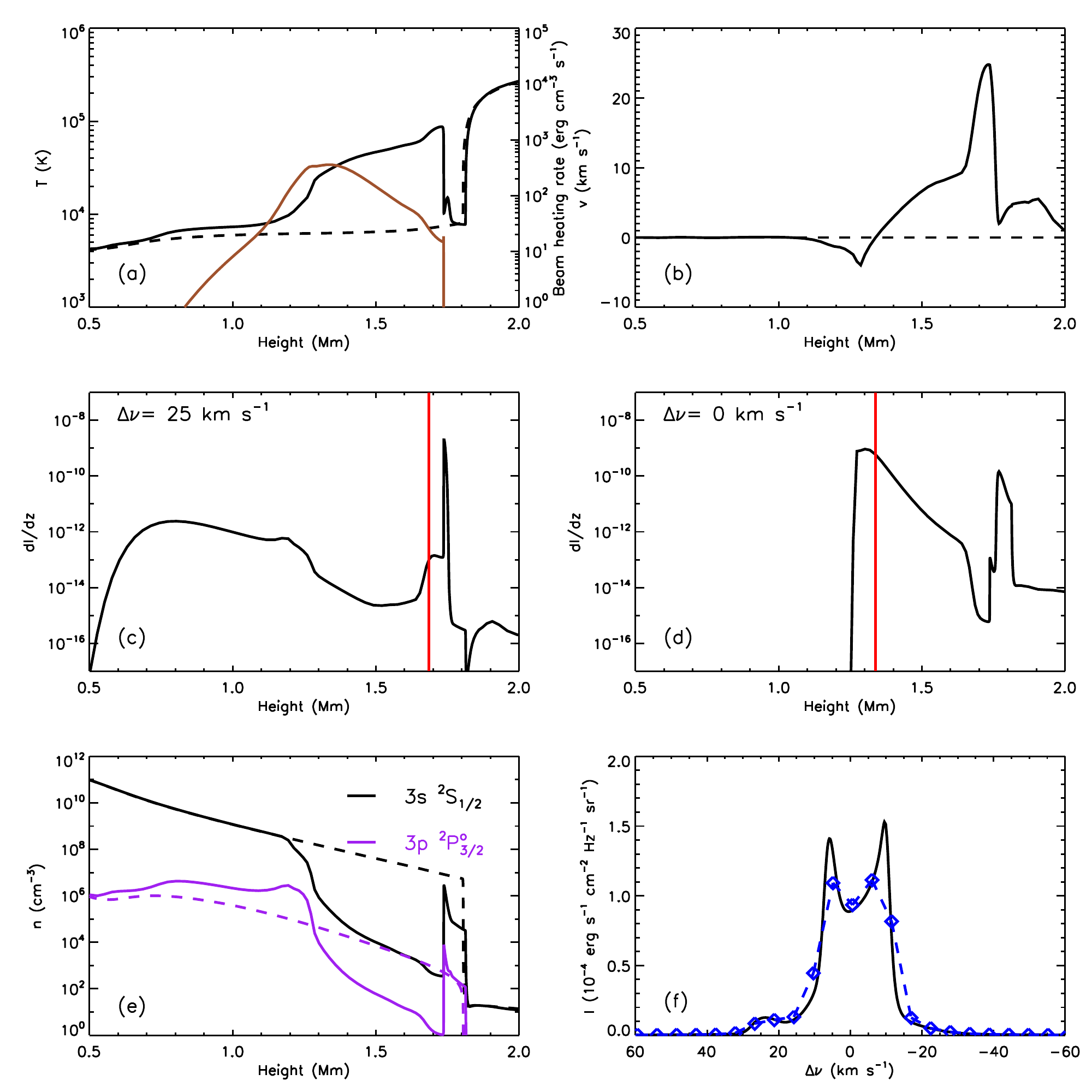}
\caption{Same as Figure~\ref{104}, but for the f10E20d7 model at 3.0 s. Note that the contribution functions in panels (c) and (d) are for $\Delta\nu = 25$ km s $^{-1}$ and $\Delta\nu = 0$ km s $^{-1}$, respectively.}
\label{107}
\end{figure}

\begin{figure}
\plotone{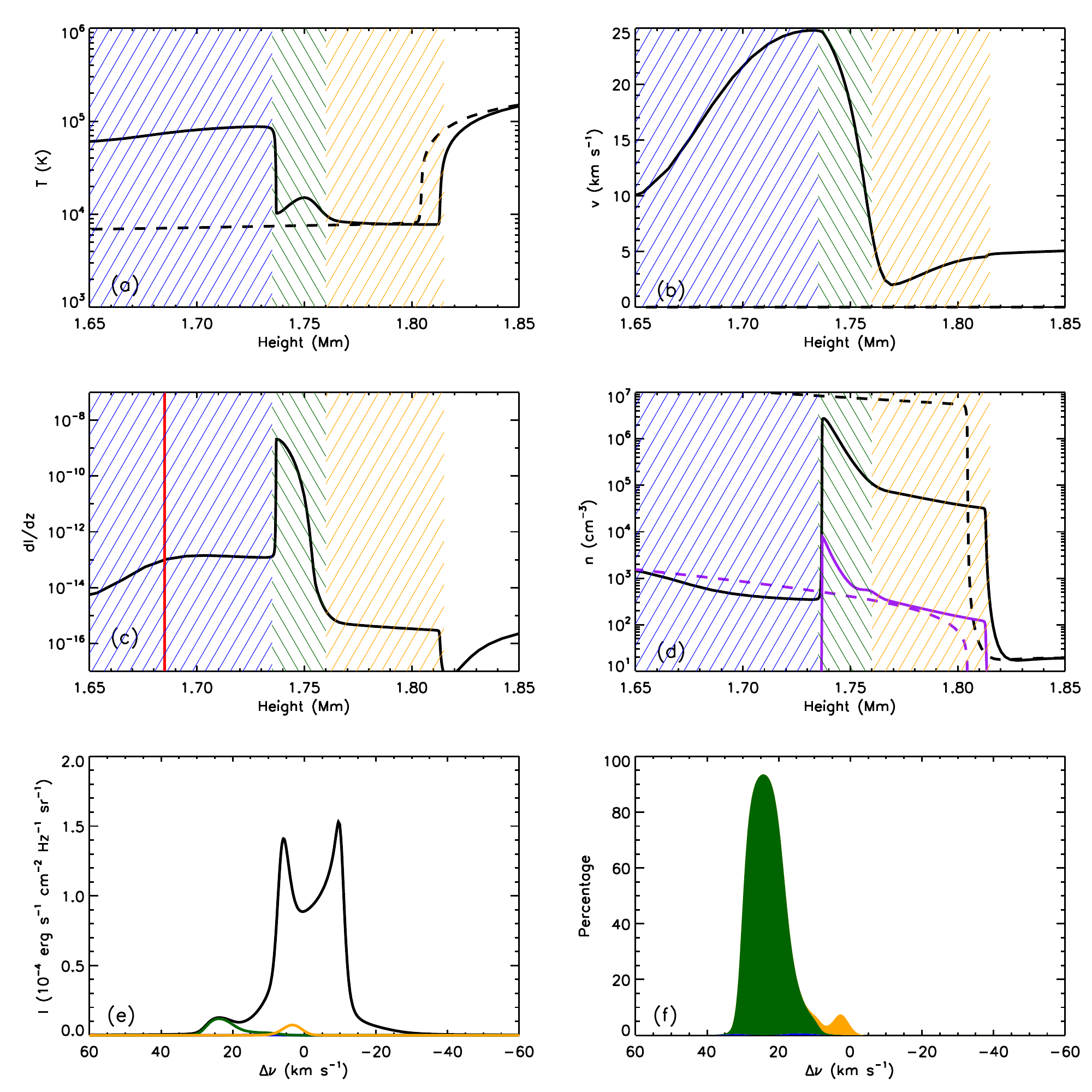}
\caption{Same as Figure~\ref{detail1}, but for the f10E20d7 model at 3.0 s. Note that the contribution function in panel (c) is for $\Delta\nu = 25$ km s $^{-1}$. The blue, green and orange lines in panel (e) and areas in panel (f) represent the values and percentage of line intensity contributed by the region that is shaded in the same color in panels (a)--(d).}
\label{detail3}
\end{figure}

\begin{figure}
\plotone{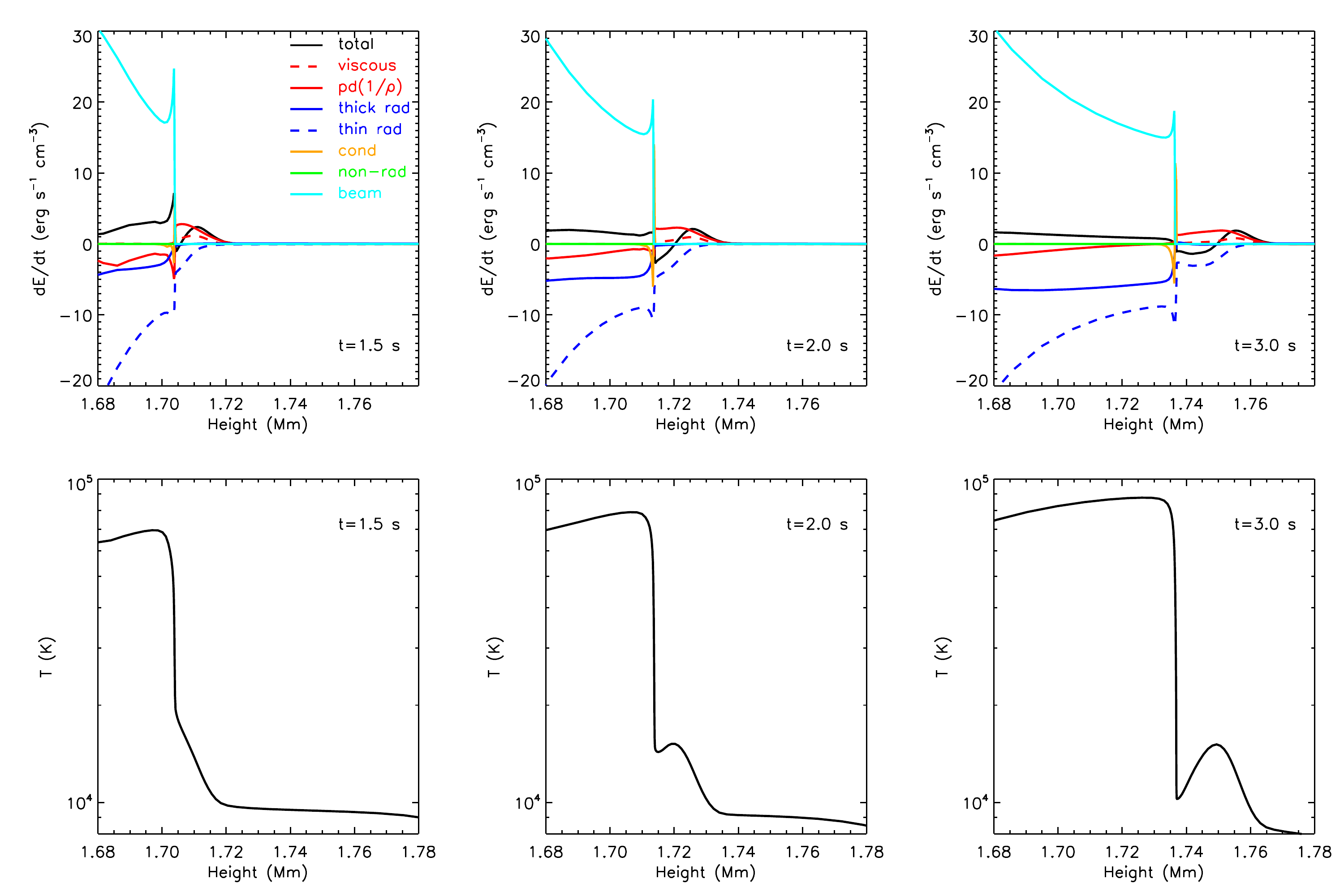}
\caption{Top row: Height distribution of different energy terms for the f10E20d7 model at 1.5 s, 2.0 s, and 3.0 s. Different terms in the energy equation are separated, including viscous heating (red dashed), work done by pressure gradient (red solid), radiative losses from thick lines (blue solid), thin radiative losses (blue dashed), thermal conduction (orange), background non-radiative heating (green) and electron beam heating (cyan). Bottom row: Height distribution of temperature for the same model as above. }
\label{blob}
\end{figure}

In observations, the \ion{Mg}{2} k blue-wing enhancement only appears in some rare events, and the observed enhancement seems less strong than what is predicted  in our simulations. Typical examples have shown only a blue-wing  asymmetry in the line profile \citep{2018tei,2019huang}. 
We have degraded the synthetic line profiles to IRIS resolution of 52 m\r{A} pixel$^{-1}$ (Figs.~\ref{104}(f), \ref{97}(f) and \ref{107}(f)), and find that the blue component are still visible, although the features in the line core are slightly smeared out.
Such a discrepancy between the observed and simulated \ion{Mg}{2} k line profiles could be mitigated by adopting larger turbulent velocities. If we increase the micro-turbulent velocities in our models, the blue peak in the line would be smoothed out and shown as a blue asymmetry instead. 
In addition, an increased Stark broadening would also help remove the discrepancy significantly \citep{2019zhu}.

\acknowledgments
We thank the referee for detailed and constructive suggestions that helped improve the paper. This work was supported by NSFC under grants 11903020, 11733003, 11873095, 11533005 and 11961131002, and NKBRSF under grant 2014CB744203. Y.L. is supported by CAS Pioneer Hundred Talents Program and XDA15052200, XDA15320301 and XDA15320103-03.

\clearpage

\end{document}